\documentclass[aps,prl,twocolumn,superscriptaddress,showpacs,amsmath,amssymb]{revtex4-1}
\usepackage{amsmath}
\usepackage{amsfonts}
\usepackage{amssymb}
\usepackage{graphicx}
\usepackage{color}

\usepackage{bm}
\usepackage{braket}
\usepackage{gensymb}

\usepackage{hyperref}

\begin{document}

\title{Designing Kitaev spin liquids in metal-organic frameworks}

\author{Masahiko G. Yamada}
\affiliation{Institute for Solid State Physics, University of Tokyo, Kashiwa 277-8581, Japan.}
\author{Hiroyuki Fujita}
\affiliation{Institute for Solid State Physics, University of Tokyo, Kashiwa 277-8581, Japan.}
\author{Masaki Oshikawa}
\affiliation{Institute for Solid State Physics, University of Tokyo, Kashiwa 277-8581, Japan.}

\date{\today}

\begin{abstract}
Kitaev's honeycomb lattice spin model is a
remarkable exactly solvable model, which has a particular type of spin
liquid (Kitaev spin liquid) as the ground state.  Although its possible
realization in iridates and $\alpha$-RuCl$_3$ has been vigorously
discussed recently, these materials have substantial non-Kitaev direct
exchange interactions and do not have a spin liquid ground state.
We propose metal-organic frameworks (MOFs) with Ru$^{3+}$ (or Os$^{3+}$)
forming the honeycomb lattice as promising
candidates for a more ideal realization of Kitaev-type spin models where
the direct exchange interaction is strongly suppressed.
The great flexibility of MOFs allows generalization to other
three-dimensional lattices, for potential realization of a variety of spin liquids
such as a Weyl spin liquid.
\\ \\
PhySH: Frustrated magnetism, Spin liquid, Quantum spin liquid, Organic compounds
\end{abstract}

\maketitle

\textit{Introduction}. --- 
Quantum spin liquids, purported exotic states of quantum magnets where
long-range magnetic orders are destroyed by quantum fluctuations,
have been a central subject in quantum
magnetism~\cite{Balents}.
As an important theoretical breakthrough,
Kitaev constructed a spin-1/2 model
on the honeycomb lattice~\cite{Kitaev} with Ising interactions
between spin components depending on bond orientations.
Its exact solution demonstrates many intriguing properties
such as fractionalized anyonic excitations.
This model was later generalized to other lattices, including
three-dimensional ones, still retaining the exact solvability~\cite{3d}.
In this paper, we call this type of model including various generalizations
as Kitaev model, and its ground states as Kitaev spin liquids.

Jackeli and Khaliullin~\cite{Jackeli} discovered that the ``Kitaev
interaction'', namely bond-dependent Ising couplings, can be realized in
a (111) honeycomb layer of iridates, i.e. the $A_2$IrO$_3$ ($A$ = Na,
Li) structure, by the superexchange interaction through the oxygen ions
due to the strong spin-orbit coupling of Ir$^{4+}$ in the Mott insulator
limit (see also Ref.~\cite{Shitade} for the itinerant limit).

However, unfortunately, it turned out that iridates and related
inorganic compounds, such as $\alpha$-RuCl$_3$~\cite{rucl}, exhibit a
conventional magnetic order at low enough temperatures and do not have a
true spin liquid ground state.  This is due to the non-Kitaev
interactions, such as antiferromagnetic Heisenberg interaction, mainly
coming from the direct exchange interaction between the metal
ions~\cite{Yamaji}.  While their finite-temperature properties still
reflect the proximity to the Kitaev model~\cite{Kimchi} and thus are of
great interest, the current situation calls for a more ideal realization
of the Kitaev model in real materials, so that they exhibit spin liquid
ground states.

In this Letter, we propose such a possible realization of the Kitaev
model in metal-organic frameworks (MOFs), crystalline materials
consisting of metal ions and bridging organic ligands.  Although MOF is
a central subject in modern complex chemistry, MOFs have not attracted
much attention in the context of magnetism.  This is perhaps because
they do not show any magnetic ordering at room temperature as direct
exchange interactions between magnetic metal ions are suppressed and the
remaining indirect superexchange interactions via nonmagnetic organic
ligands are weak.  We take the advantage of this suppression of direct
exchange interactions to realize the Jackeli-Khaliullin mechanism,
i.e. superexchange realization of the Kitaev interaction.
Furthermore, 
based on tight-binding models and the fragment molecular orbital (fMO)
method~\cite{Kitaura} in combination with the density functional
theory (DFT) calculations, we demonstrate that
the Jackeli-Khaliullin mechanism gives rise to the dominant
Kitaev interactions with oxalate-based (or tetraaminopyrazine-based) ligands.
This opens up the possibility of \emph{designing} the appropriate
MOFs to realize Kitaev spin liquids.

\begin{figure}
\centering
\includegraphics[width=7cm]{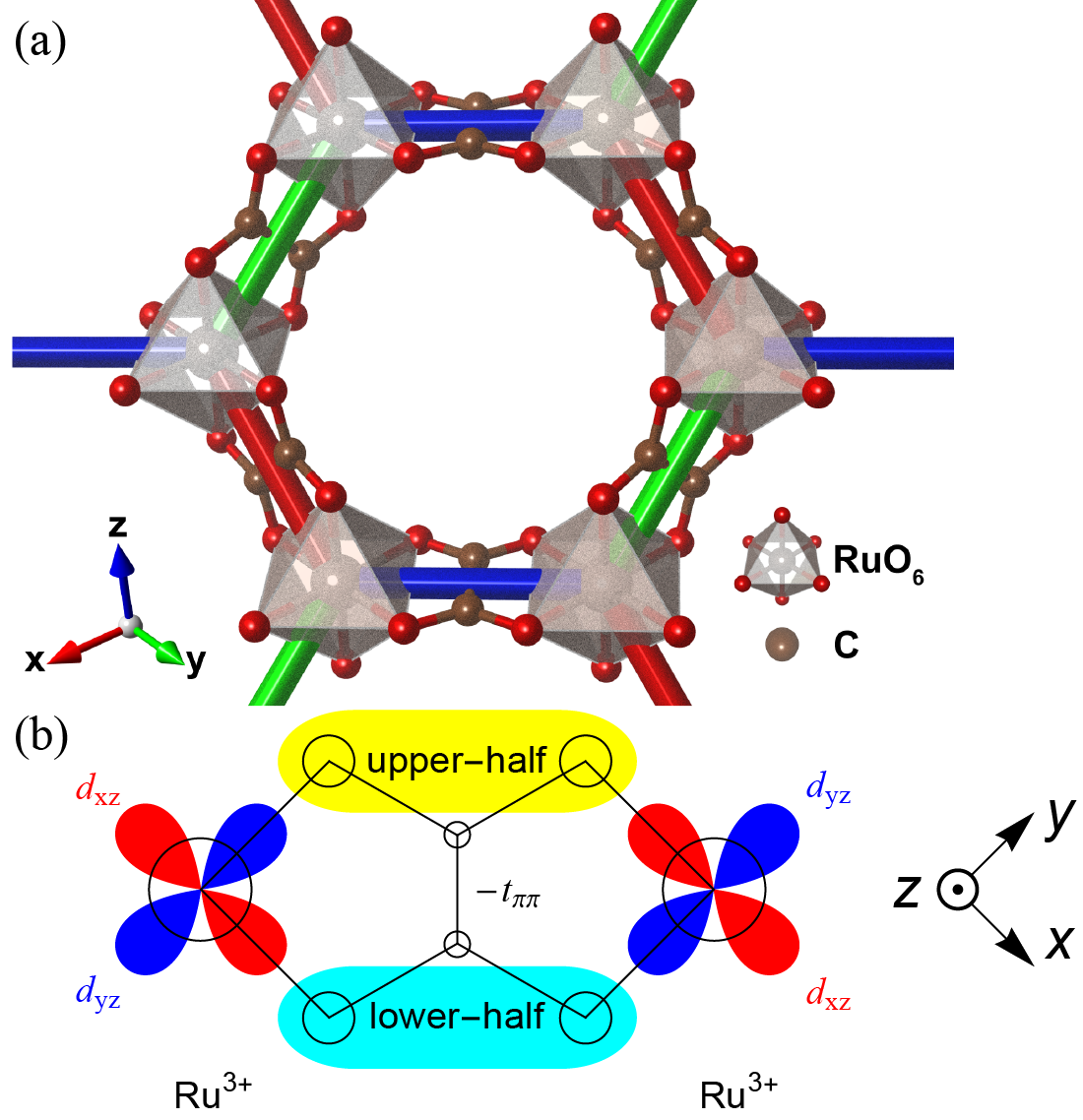}
\caption{(Color online) (a) Geometric structure of honeycomb Ru-oxalate frameworks. White octahedra are RuO$_6$ octahedra and carbon atoms are shown in brown. The color of the bond between the Ru atoms means which plane the bridging oxalate belongs to (red: $yz$-plane, green: $zx$-plane, blue: $xy$-plane). (b) Two superexchange pathways between two neighboring Ru$^{3+}$ through an oxalate ion belonging to the $xy$-plane (one of the blue bonds in (a)). $t_{\pi\pi}$ is a hopping parameter between the two distinct pathways.
}
\label{str}
\end{figure}

\textit{Structures of the Proposed Metal-Organic Frameworks}. --- 
In order to realize a Kitaev spin liquid in MOFs using the
Jackeli-Khaliullin mechanism~\cite{Jackeli}, we first propose an MOF
structure with Ru$^{3+}$ (or Re$^{2+},$
Os$^{3+},$ Rh$^{4+},$ Ir$^{4+}$) ions in the octahedral
coordination.  Because of the composite effects of the octahedral ligand
field and the strong spin-orbit coupling, these 4$d^5$ or 5$d^5$ ions
show a low-spin ground state with an effective angular momentum
$J_{\textrm{eff}}=1/2.$
Hinted by the (111) honeycomb layers of iridates,
we propose a geometric structure shown in Fig.~\ref{str}(a),
where the RuO$_6$ octahedra form a two-dimensional (2D) honeycomb lattice
and the organic ligand (in this case oxalate, (C$_2$O$_4$)$^{2-},$ or ox$^{2-}$)
connects the two edges of the octahedra.
Indeed, many honeycomb MOFs with this structure have already been found by chemists~\cite{Weiss,Mathoniere,Abrahams,Coronado,Abrahams2,Luo,Shilov2012,Atzori,Abherve,Zhu,Kitagawa,Harris}.
More interestingly, this honeycomb structure of metal-oxalate frameworks
is also found in nature in the form of minerals, stepanovite and
zhemchuzhnikovite~\cite{Huskic}.  Thus, we can expect that
this honeycomb geometry is chemically stable.
Moreover, experiments~\cite{Fujino} for the molecule
[\{Ru(acac)$_2$\}$_2$($\mu$-ox)]$^-$ (acac$^-$ = acetylacetonate) observed
an anisotropic spin interaction via oxalate due to the spin-orbit coupling
of Ru$^{3+}$ with electron spin resonance.  Similar anisotropy has also been
observed in many other Ru and Os complexes~\cite{Kaim}.
It is, therefore,
natural to consider Ru and oxalate to realize the Kitaev interaction,
as the first candidate.

\begin{figure}
\centering
\includegraphics[width=8.6cm]{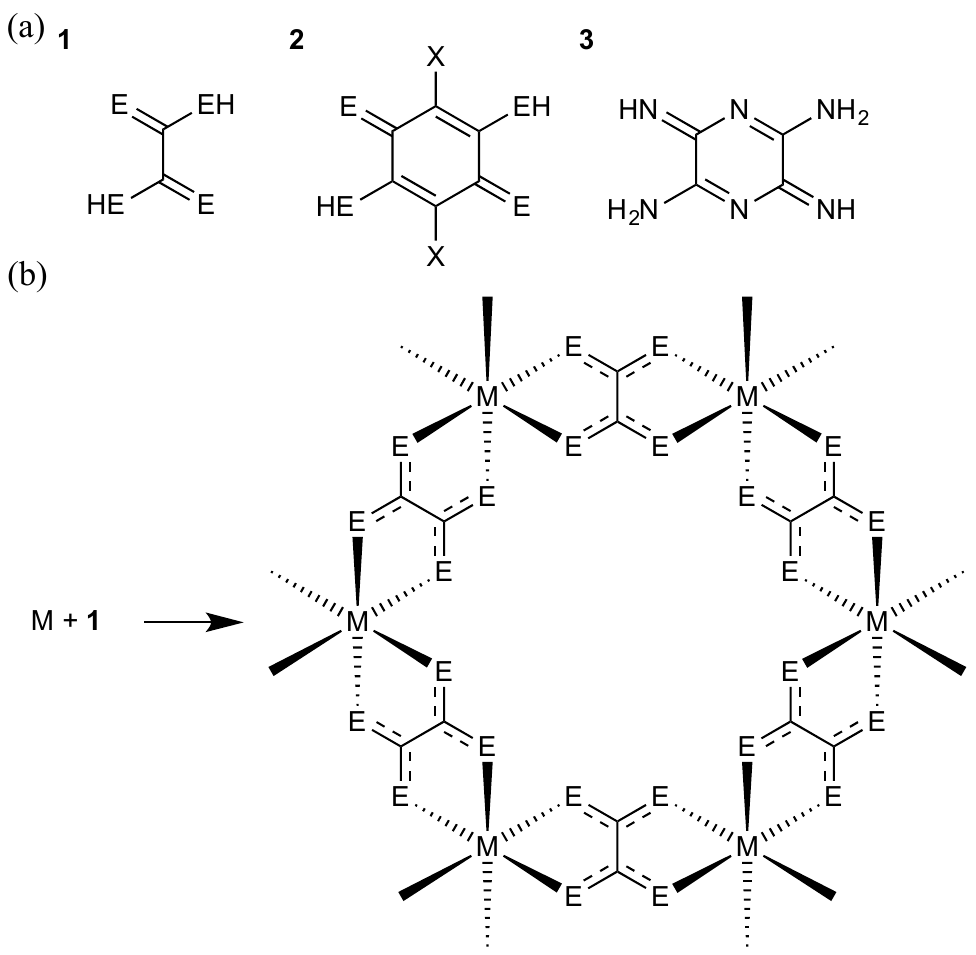} 
\caption{Chemical formulae for the proposed metal-organic frameworks.
(a) Possible organic molecules to realize a honeycomb structure with octahedral coordination. \textbf{1}: Oxalate-based molecules ($E$ = O, S, NH). \textbf{2}: Quinoid-based molecules ($E$ = O, S, NH; $X$ = H, Cl, Br, I, etc.). \textbf{3}: Tetraaminopyrazine-based molecule. (b) Honeycomb structure of metal-oxalate frameworks ($M$ = Ru, Os).
}
\label{chem}
\end{figure}

The ligand can be replaced with other organic molecules to achieve
a wide variety of MOFs.
Some possibilities, including a newly proposed one, are listed in Fig.~\ref{chem}(a) for MOFs
with Ru$^{3+}$ (or Os$^{3+}$) in the octahedral coordination.
In Fig.~\ref{chem}(a), \textbf{1} is oxalic acid ($E$ = O) and becomes oxalate
in the proposed MOF (see Fig.~\ref{chem}(b)).  In the case of $E$ = S (resp. NH),
we call it tetrathiooxalate (resp. tetraaminooxalate).  Similarly,
\textbf{2} becomes dhbq$^{2-}$ ($E$ = O, $X$ = H and dhbq = 2,5-dihydroxy-1,4-benzoquinone)
or $X_2$An$^{2-}$ ($E$ = O, $X$ = Cl, Br, etc. and An = anilate), and
\textbf{3} is tetraaminopyrazine C$_4$N$_6$H$_6$ and becomes (C$_4$N$_6$H$_4$)$^{2-},$
which we have newly proposed.
There already exists a metal-oxalate framework including Ru$^{3+},$
such as LaRu(ox)$_3 \cdot$10H$_2$O in Ref.~\cite{Kitagawa}, and
the molecule [Ru(ox)$_3$]$^{3-}$ is known to be a good spin-1/2 qubit~\cite{Freedman},
so it is very natural to use this Ru(ox)$_3$ unit as a building block
for highly entangled quantum states.
In any $M_2L_3$ ($M$ = Ru, Os, and $L$ = ox, dhbq, etc.) structures, the metal ion $M$
should be in the $3+$ state and the organic ligand $L$ should be in the $2-$ state.
Additional structures may be necessary to maintain the rigid honeycomb
structure for $M_2L_3$ layers, but it will not affect the effective spin model
as long as Ru or Os is in the $3+$ state and the interlayer interaction is negligible.
In fact, in the Fe$^{3+}$-based MOF with layered structure discovered in
Ref.~\cite{Harris}, the interlayer distance of metal ions is as large as
8.7449 \AA, and the interlayer interaction is found to be negligible or
ferromagnetic.

\textit{Superexchange Interaction}. --- 
The main obstacle to realize Kitaev spin liquids
in inorganic materials was the direct exchange interaction between the
metal ions, which yields significant non-Kitaev interactions~\cite{Yamaji}.
In MOFs, the electron density of the bridging organic ligand screens the
wavefunction tails of the metal ions, which would substantially reduce the
direct overlap between orbitals of the neighboring metals.  Thanks to
this, the direct exchange interaction is strongly suppressed in most MOFs~\cite{Harris}.
This is the most important advantage of using organic ligands
for the realization of Kitaev spin liquids compared to other
inorganic candidates, such as iridates and $\alpha$-RuCl$_3$.
However, non-Kitaev interactions can also arise from the
superexchange interaction.
In order to evaluate the magnitude of those terms,
we derive the effective spin Hamiltonian
in the following steps.
First we obtain the effective tight-binding model for the superexchange
between two Ru ions, and then we map this model into the effective spin model.
As a concrete example, we
take one specific bond belonging to the $xy$-plane shown
in Fig.~\ref{str}(b) in the following discussions.

As we will see below, Jackeli-Khaliullin mechanism works perfectly
to give rise to the pure Kitaev interaction,
if there are two separate superexchange pathways connecting
$d_{xz}$ and $d_{yz}$ only.
The advantage of using oxalate as the ligand is the existence of
the localized modes along upper- and lower- edges of the molecule,
which are analogous to the localized edge modes along the zigzag edges of
graphene~\cite{Fujita}.
They can function as the desired two superexchange pathways.
In fact, the separation of the two paths is not perfect and we
have to consider the effect of their mixing quantitatively.
Nevertheless, as we will demonstrate, the mixing is small and
the two superexchange pathways are approximately protected.
This leads to a dominant Kitaev interaction.

In the chemical terminology, we may consider two ``fragment molecular
orbitals'' (fMO)~\cite{Kitaura} corresponding to these localized edge
modes.
For simplicity, here we focus on
HOMOs (highest occupied molecular orbitals) and LUMOs
(lowest unoccupied molecular orbitals)
coming from $\pi$-orbitals
($\pi$-HOMO and $\pi^*$-LUMO for short, respectively),
which are the most important as superexchange pathways between
Ru$^{3+}$ ions.
With the annihilation operators $u$ and $l,$ respectively, for $\pi$-HOMOs along
upper- and lower-edges in Fig.~\ref{str}(b), the model Hamiltonian may be written as
\begin{align}
        H_{d\pi} & = -t_{d\pi} (u^\dagger d_{yz}^L+l^\dagger d_{xz}^L+u^\dagger d_{xz}^R+l^\dagger d_{yz}^R+h.c.)
\notag \\
& +V_\pi(u^\dagger u+l^\dagger l)-t_{\pi\pi} (u^\dagger l+l^\dagger u),
\label{eq.fMO}
\end{align}
where
$t_{d\pi}$ is the hopping matrix element between the Ru
$t_{2g}$-orbitals and the $\pi$-HOMOs,
$V_\pi$ is the energy level of $\pi$-HOMOs,
$t_{\pi\pi}$ is the tunneling matrix element between the two fMOs,
$d_{i}^L$ (resp. $d_{i}^R$) is the annihilation operator of an electron
on the Ru $d_i$-orbital on the left (resp. right) side in
Fig.~\ref{str}(b).
Similar terms exist for the $\pi^*$-LUMO with the energy level
$V_{\pi^*}$ or the hopping and tunneling matrix elements
$t_{d\pi^*}$ and $t_{\pi^*\pi^*},$ respectively.
If the separation of the two paths were perfect, then $t_{\pi\pi}=0,$
but in reality it is non-vanishing.
Thus the energy levels of HOMOs are split into $V_\pi\pm t_{\pi\pi}$.
Nevertheless, reflecting the approximate protection of the two
pathways, $t_{\pi\pi}$ is relatively small.
This can be confirmed by the DFT calculations,
as we will discuss later.

Integrating over the $\pi$-HOMO states $u$ and $l,$ we obtain the
effective hopping terms between the two Ru$^{3+}$ ions as
\begin{align}
        H_{dd} =& -t_1 (d_{yz}^{L\dagger} d_{yz}^R+d_{xz}^{L\dagger} d_{xz}^R)-t_2 (d_{yz}^{L\dagger} d_{xz}^R+d_{xz}^{L\dagger} d_{yz}^R) \nonumber \\
        &-t_3 d_{xy}^{L\dagger} d_{xy}^R +h.c. ,
\label{eq.Hdd}
\end{align}
where 
$t_1 = t^2_{d\pi}t_{\pi\pi}/(V^2_\pi-t^2_{\pi\pi})$,
$t_2 = t^2_{d\pi}V_{\pi}/(V^2_\pi-t^2_{\pi\pi})$, and $t_3=0$.
As expected, $t_1 / t_2$ is small when $t_{\pi\pi}$ is small.
We should also include the superexchange contributions
through $\pi^*$-LUMO and $\sigma$-orbitals in deriving the effective
hopping terms.

Once we obtain the effective hopping~\eqref{eq.Hdd},
by projecting onto $J_\textrm{eff}=1/2$ states of Ru$^{3+}$,
we can derive the effective spin model~\cite{JKG2}
\begin{equation}
        H=\sum_{\langle ij\rangle \in \alpha \beta(\gamma)} [J \bm{S}_i \cdot\bm{S}_j +KS_i^\gamma S_j^\gamma +\Gamma (S_i^\alpha S_j^\beta +S_i^\beta S_j^\alpha)], \label{eq.JKGamma}
\end{equation}
where $J$ is the Heisenberg coupling, $K$ is the Kitaev coupling, and $\Gamma$ is the symmetric off-diagonal exchange. The explicit form of these parameters is included in
Supplemental Material. $\alpha,\beta,\gamma \in \{x,y,z\}$
and $\langle ij\rangle \in \alpha \beta(\gamma)$ 
mean that the bond plane of the nearest-neighbor bond $\langle ij\rangle$
is the $\alpha\beta$-plane perpendicular to the $\gamma$-axis.
This model is an extended version of the Kitaev-Heisenberg
model~\cite{Chaloupka}, known as the JK$\Gamma$ model~\cite{JKG1,JKG2}.
In the limit $J/|K| \to 0, \Gamma/|K| \to 0$, the model is nothing but
the honeycomb Kitaev model~\cite{Kitaev}, which has a gapless spin
liquid ground state.
Here we ignore Dzyaloshinskii-Moriya interactions,
assuming the parity symmetry around the bond center.

In this way, we can estimate the parameters $J$, $K$, and $\Gamma$
starting from the fMO-based model~\eqref{eq.fMO} of the ligand.  As we
have emphasized, if the superexchange paths along the upper- and lower-
halves were completely separate, $t_{\pi\pi}=t_{\pi^*\pi^*}=0$ which
would give $t_1=t_3=0$.  With the effective hopping~\eqref{eq.Hdd} with
only $t_2$ non-vanishing, Jackeli-Khaliullin mechanism works perfectly
and we would obtain the ideal Kitaev model with $J=\Gamma=0$.
This condition could easily be
met by using two formates as bridging ions, each of which acts as
a separate superexchange pathway, although the honeycomb structure may be
unstable in metal-formate frameworks.

\textit{Estimation of Spin Interactions}. --- 
In order to estimate the parameters in the effective spin
model~\eqref{eq.JKGamma} in real MOFs proposed in Fig.~\ref{chem}, we have performed DFT calculations for
the oxalate ligand using 
\textsc{openmx}~\cite{openmx} software package.
It should be noted that a calculation on a ligand molecule
only gives the energy differences such as $V_{\pi^*}-V_\pi.$
The individual energy levels such as $V_{\pi^*}$ and $V_\pi$
measured from the Fermi level, i.e. the Ru $J_\textrm{eff}=1/2$ orbital,
cannot be directly obtained.

In this work, as a crude but quick estimate to see the potential of our
proposal, we will proceed as follows.  In the case of oxalate, for
example, $V_{\pi^*}-V_\pi = 6.47$ eV from DFT.  Experiments~\cite{Fujino} suggest
that the metal to ligand charge transfer (MLCT) energy
$E_{\mbox{\scriptsize LUMO}}=V_{\pi^*}-t_{\pi^*\pi^*} \sim 2.6$ eV,
which corresponds to the optical absorption at the wavelength of 485 nm.
This, together with $t_{\pi\pi}=0.153$ eV and
$t_{\pi^*\pi^*}=1.631$ eV from the DFT calculations, implies $V_{\pi^*}
\sim 4.2$ eV and $V_\pi \sim -2.3$ eV.  Using these parameters,
and $t_{d\pi^*}/t_{d\pi}\sim 0.6159$ for oxalate,
we find the ratio between the effective hoppings in Eq.~\eqref{eq.Hdd} as
\begin{equation}
\frac{t_1}{t_2} = \frac{\frac{t_{d\pi}^2 t_{\pi\pi}}{V_\pi^2-t_{\pi\pi}^2}-\frac{t_{d\pi^*}^2 t_{\pi^*\pi^*}}{V_{\pi^*}^2-t_{\pi^*\pi^*}^2}}{\frac{t_{d\pi}^2 V_\pi}{V_\pi^2-t_{\pi\pi}^2}-\frac{t_{d\pi^*}^2 V_{\pi^*}}{V_{\pi^*}^2-t_{\pi^*\pi^*}^2}} \sim 0.023.
\end{equation}
Taking the superexchange via $\sigma$-orbitals into account in
a similar manner, we find $t_3/t_2 \sim -0.196$.
From these values, we estimate $J/|K| \sim 0.004$ and
$|\Gamma|/|K| \sim 0.15,$ namely the Kitaev interaction is
strongly dominant.
The details of the derivation is given in Supplemental Material.

In general, we find
that the resulting low-energy effective
model is dominated by the Kitaev interaction
($J/|K| \sim |\Gamma|/|K| \lesssim 1/10$),
if the conditions
$|t_{\pi\pi}|/|V_\pi| \lesssim 1/10$ and
$|V_\pi|/|V_{\pi^*}| \lesssim |t_{d\pi}\sqrt{t_{\pi\pi}}|/|t_{d\pi^*}\sqrt{t_{\pi^*\pi^*}}|$
are both met.
We note that the smallness of $|V_\pi|/|V_{\pi^*}|$
implies that the superexchange is hole-mediated~\cite{Browne}.

Although there is no particular reason to have degeneracy in aromatic
ligands, such as dhbq$^{2-}$ and $X_2$An$^{2-}$ (\textbf{2} in
Fig.~\ref{chem}(a)),
it is possible to have similar degeneracy of
$\pi$-HOMOs in the tetraaminopyrazine-based ligand
(C$_4$N$_6$H$_4$)$^{2-}$ (\textbf{3} in
Fig.~\ref{chem}(a)).
In contrast to oxalate, the degenerate two $\pi$-HOMOs are just below
the Fermi energy even in the vacuum, so tetraaminopyrazine-based
structures should also be good candidates for Kitaev-dominant MOFs.  In
addition, this (C$_4$N$_6$H$_4$)$^{2-}$ would stabilize the planar
structure more easily than oxalate due to the $\pi$-conjugated
nature.

While the present estimate is crude and we have ignored many
possible corrections, these results suggest that our proposal
of realizing Kitaev spin liquids in MOFs is quite promising.
We emphasize that, MOFs have the flexibility in the choice of
ligand molecules, so that many possibilities can be tried
for the realization of Kitaev spin liquids.

\textit{Designing a Variety of Kitaev Spin Liquids}. ---
Here we emphasize another advantage of MOFs: we can construct various
complex geometric structures, not limited to the honeycomb lattice, by
self-organization.  In particular, three-dimensional (3D) generalizations of the Kitaev
model are of great interest~\cite{Mandal}.  A realization in iridates
has been reported but again with a magnetic ordering at low
temperatures~\cite{Takayama}.  One of the 3D structures known as the
hyperhoneycomb lattice or (10,3)-$b$ is, on the other hand, naturally
realizable in MOFs, such as
[(C$_2$H$_5$)$_3$NH]$_2$Cu$_2$(C$_2$O$_4$)$_3$ shown in Fig.~S1 in
Supplemental Material, and can in principle be constructed just by
putting building blocks altogether and stirring~\cite{hyper}.  The
cation-templating is known to be important to construct a 3D
structure~\cite{hyper}, so we would possibly need to replace Cu$^{2+}$
with Re$^{2+}$ rather than with Ru$^{3+}$.
The analysis in the previous section can also be applied to MOFs with 3D
tricoordinated lattices,
to derive the same JK$\Gamma$ model as the effective low-energy
spin model on the corresponding lattice.

By applying a magnetic field to break the time-reversal symmetry, the
system with a 3D hyperhoneycomb lattice is expected to show a gapless
Weyl spin liquid ground state~\cite{Hermanns}.  More interestingly, there
are other 3D tricoordinated lattices with exotic Majorana states, which
have not been found in iridates but are possible in 3D MOFs.  Among these, the
hyperoctagon lattice or
(10,3)-$a$ structure~\cite{Wells,Tamaki,Decurtins,Coronado2,Clement,anilateoct,quinoid,Dikhtiarenko}
has Majorana Fermi surfaces, which would be destabilized by an
additional time-reversal interaction leaving an odd number of nodal
lines~\cite{Maria2014,Hermanns2}.

Finally, we would like to discuss the possibility to realize gapped spin
liquid ground states, i.e. topological phases with ground state
degeneracy.  Kitaev~\cite{Kitaev} pointed out that a non-Abelian gapped
topological phase would emerge from the honeycomb Kitaev model by
applying a magnetic field along the [111] direction in Fig.~\ref{str}(a),
and an Abelian gapped $Z_2$ topological phase, described by the toric
code~\cite{toriccode}, can be induced by introducing the bond
anisotropy, i.e. breaking the three-fold symmetry of the system. The
former situation is simply expected when we apply an external magnetic
field to the proposed honeycomb MOFs in the same way as iridates.
Concerning the latter possibility of the gapped
Abelian $Z_2$ topological phase,
realization in iridates would require an application of a uniaxial strain,
which turns out to be experimentally difficult.
In contrast, in MOFs, the bond anisotropy may be introduced
chemically using heterogeneous organic ligands, leading
to the $Z_2$ topological phase, as we present details in Supplemental Material.
An example of possible MOF structures with heterogeneous ligands
is shown in Fig.~S6.
Similar distorted honeycomb MOFs with heterogeneous ligands were already
reported~\cite{alternating}, so we expect that the materials proposed
here could be synthesized.

\textit{Conclusion}. --- 
We discussed the possibility to realize Kitaev spin liquids in MOFs and
found three advantages over inorganic materials: the suppression of
undesired direct exchange interactions,
the flexibility to control the parameters using a variety of
possible ligands,
and the natural realization of
complex structures.
All of these features of MOFs will pave a way towards an experimental
realization of the exotic Kitaev spin liquids as the ground states, one
of the holy grails in contemporary condensed matter physics.

Due to the large unit cell of MOFs, similarly to new
Kitaev-dominant iridates with a longer Ir --- O --- O --- Ir
superexchange~\cite{IrOOIr}, the energy scale of the superexchange
interaction will be 10--100 K.
The finite-temperature phase transition into 
the Kitaev spin liquid phase in the 3D case is expected at
$1/100$ of this energy scale~\cite{Nasu}, namely at 0.1--1 K. Although
this will make the experimental studies of the Kitaev spin liquid phase below
this temperature somewhat challenging, it is still possible as it is
reachable with a dilution refrigerator.

Our proposal opens up many questions.
We have not discussed the geometric stability of the proposed MOFs,
but there is no obvious obstacle for realization of
metal-oxalate frameworks with Ru$^{3+}$ considering
many reports of synthesizing metal-oxalate frameworks with various
metals~\cite{Mathoniere,Coronado,Zhu,Kitagawa,hyper,Tamaki,Decurtins,Coronado2,Clement,Dikhtiarenko}.
In fact, honeycomb MOFs with $R3c$ and $P6/mmm$ space groups,
which respect a full symmetry of the honeycomb lattice, was realized with the high-spin
Fe$^{3+}$ ions~\cite{Mathoniere}.
We can thus naturally expect that replacement
by the low-spin Ru$^{3+}$ would not lead to any lattice distortions from
the idealized case~\footnote{See Supplemental Material at [URL will be inserted by
publisher] for the lattice distortions in 3D structures.} because of the proximity of the ionic radii,
in contrast to honeycomb iridates with small monoclinic distortions~\cite{JKG2}.
It is still subtle whether a possible trigonal distortion~\cite{trigonal},
which does not break any space group symmetry, supports
the formation of our proposed structure to stabilize the spin liquid
states~\cite{Ohrstrom,Gruselle}. Assuming the effect of the trigonal distortion
of the RuO$_6$ octahedra in MOFs is the same as that in iridates, a recent study~\cite{JKG2}
discovered that the Kitaev term would dominate when $\angle$ (O --- Ru --- O) $\sim 80\degree,$
which quite agrees with the observed values in MOFs~\footnote{For example, the observed
bond angles in various MOFs discussed in Ref.~\cite{Dikhtiarenko} range from
81\degree~to 83\degree, which allows us to fine-tune the parameters by the chemical control}.
In any case, we will need a more rigorous first-principles calculation with geometric
optimization.

\begin{acknowledgments}
We thank A.~Banisafar, L.~E.~Darago, D.~E.~Freedman, T.~D.~Harris, M.~Hermanns, G.~Jackeli and J.~R. Long in particular for illuminating discussions. We also wish to acknowledge H.~Aoki, M.~Dinc\u{a}, M.~Gohlke, T.~Hamai, D.~Hirai, H.~Katsura, I.~Kimchi, T.~Ozaki, F.~Pollmann, T.~Soejima, Y.~Tada, H.~Takagi, N.~Tsuji, S.~Tsuneyuki, H.~Watanabe and J.~Yamazaki for helpful comments. The computation in this work has been done using the facilities of the Supercomputer Center, the Institute for Solid State Physics, the University of Tokyo. The crystal data included in this work have been taken from the Cambridge Crystallographic Data Centre.
M.G.Y. is supported by the Materials Education program for the future leaders in Research, Industry, and Technology (MERIT), and by JSPS.
H.F. is supported by JSPS through Program for Leading Graduate Schools (ALPS). This work was supported by JSPS KAKENHI Grant Numbers JP15H02113, JP16J04752, JP17J05736, and by JSPS Strategic International Networks Program No. R2604 ``TopoNet''.  This research was supported in part by the National Science Foundation under Grant No. NSF PHY-1125915.
\end{acknowledgments}

\bibliography{paper}

\end{document}


\title{Supplemental Material for \\ ``Designing Kitaev spin liquids in metal-organic frameworks''
}

\author{Masahiko G. Yamada}
\affiliation{Institute for Solid State Physics, University of Tokyo, Kashiwa 277-8581, Japan.}
\author{Hiroyuki Fujita}
\affiliation{Institute for Solid State Physics, University of Tokyo, Kashiwa 277-8581, Japan.}
\author{Masaki Oshikawa}
\affiliation{Institute for Solid State Physics, University of Tokyo, Kashiwa 277-8581, Japan.}

\maketitle

\onecolumngrid

In this Supplemental Material, we have the following sections.
Section~A: the calculation details of the molecular orbitals within density functional theory (DFT),
Section~B: the calculated results for the molecular orbitals,
Section~C: the derivation of the edge states of oxalate-based ligands from the tight-binding model,
Section~D: the derivation of the JK$\Gamma$ model from the microscopic model and the order estimation of the ratio of the parameters with molecular orbital diagrams for the proposed ligands,
Section~E: the microscopic spin model for the metal-organic frameworks with heterogeneous organic ligands,
and Section~F: the lattice distortions in three-dimensional structures.

\section*{S\lowercase{ection}~A: Profiles for the DFT calculations}

Our analysis is based mostly on simple tight-binding models or
on a fragment molecular orbital method~\cite{Kitaura} in combination with a DFT method
for organic molecules. We used a first-principles electronic structure calculation code
called \textsc{openmx}~\cite{openmx} for the DFT calculations. In \textsc{openmx},
one-particle wave functions are expressed by the linear combination of pseudoatomic basis
functions (LCPAO) and the norm-conserving pseudopotentials are used. We used the
generalized gradient approximation represented by Perdew, Burke and Ernzerhof
(GGA-PBE)~\cite{PBE} for the exchange-correlation potential. A self-consistent loop
was iterated until the energy was relaxed with the error of $10^{-10}$ Hartree.
The geometric optimization of internal coordinates was also iterated until the force
became smaller than $10^{-4}$ Hartree/Bohr with a constraint on the molecules to
be completely planar.

In order to calculate the molecular orbitals of the proposed ionic ligands,
we used an energy cutoff of 500 Ry for the numerical integrations and the solution
of the Poisson equation using the fast Fourier transformation algorithm.
As basis functions, 2$s$, 2$p$, 3$s$, 3$p$, and 3$d$-orbitals
for C, N, O, 3$s$, 3$p$, 3$d$, 4$s$, and 4$p$-orbitals for S,
and 1$s$, 2$s$, 2$p$-orbitals for H were employed. In \textsc{openmx},
the excess charge of the ionic state of the ligands is compensated by
a uniform background electronic charge with an opposite sign, which is
an artificial approximation for the numerical reason. All the calculations
are done without including a spin-orbit coupling or spin polarization.
The results of these calculations are included mostly in this Supplemental Material.

\begin{figure}[H]
\centering
\includegraphics[width=8cm]{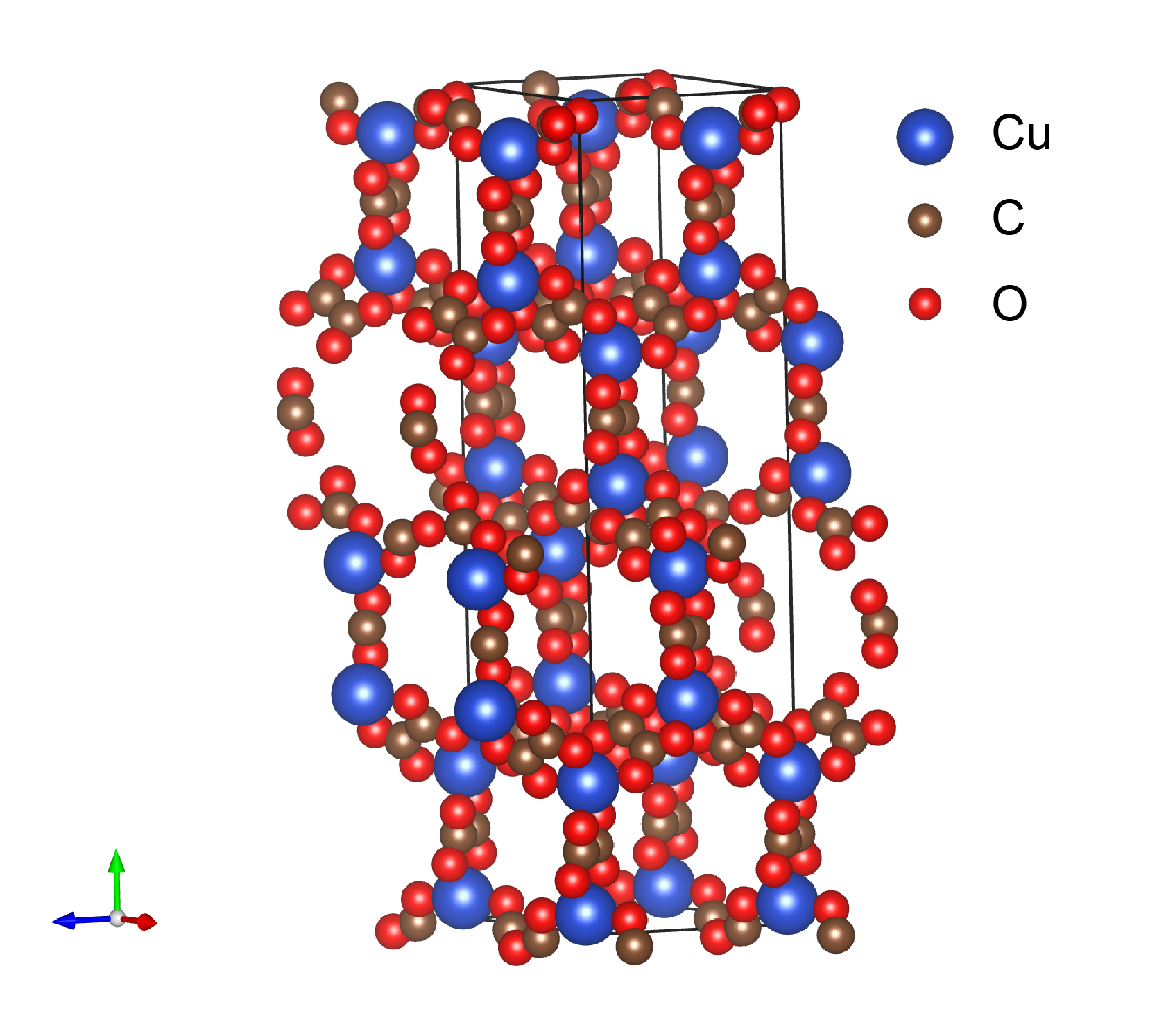}
\caption{Hyperhoneycomb structure of a metal-oxalate framework. This structure is included in~\cite{hyper} (Cu: blue, C: brown, O: red), reconstructed from the crystal data taken from CSD-KEDJAG in the Cambridge Structural Database~\cite{CSD}.
The auxiliary structure other than metal ions and oxalate ions are omitted for simplicity.}
\label{hyper}
\end{figure}

In addition, the crystal data for Fig.~S\ref{hyper} was taken from
the Cambridge Structural Database~\cite{CSD} in order to show the hyperhoneycomb structure.

\section*{S\lowercase{ection}~B: Calculated molecular orbitals}

\begin{figure}
\centering
\includegraphics[width=8cm]{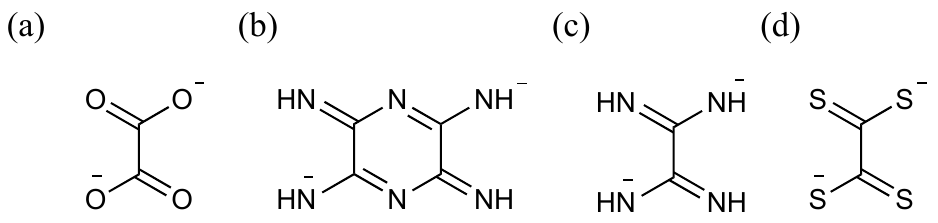}
\caption{Structure formulae of the proposed ligands listed in Table~\ref{param}.
(a) Oxalate, an ionized form of \textbf{1} with $E$ = O in Fig.~2(a) in the main text.
(b) (C$_4$N$_6$H$_4$)$^{2-}$, an ionized form of \textbf{3} in Fig.~2(a) in the main text.
(c) Tetraaminooxalate, an ionized form of \textbf{1} with $E$ = NH in Fig.~2(a) in the main text.
(d) Tetrathiooxalate, an ionized form of \textbf{1} with $E$ = S in Fig.~2(a) in the main text.
}
\label{mf}
\end{figure}

Using the conditions for \textsc{openmx}~\cite{openmx} described in Section~A,
we calculated the molecular orbitals near the Fermi energy for three oxalate-based ligands
(\textbf{1} with $E$ = O, NH, S in Fig.~2(a) in the main text, corresponding to (a), (c), (d) in Fig.~S\ref{mf}, respectively) and one
tetraaminopyrazine-based ligand (\textbf{3} in Fig.~2(a) in the main text, corresponding to (b) in Fig.~S\ref{mf}),
based on DFT. After the geometric optimization, all the molecules are relaxed
to have a $D_{2h}$ symmetry.

\begin{table}
        \centering
        \caption{\label{param}Parameters for each molecule estimated from DFT calculations.}
        \begin{ruledtabular}
        \begin{tabular}{cccccc}
                & molecule & formula & $t_{\pi\pi}$ (eV) & $t_{\pi^*\pi^*}$ (eV) & $V_{\pi^*}-V_\pi$ (eV) \\
                \hline
                (a) & oxalate & (C$_2$O$_4$)$^{2-}$ & 0.153 & 1.631 & 6.474 \\
                (b) & (C$_4$N$_6$H$_4$)$^{2-}$ & & 0.208 & 1.501 & 4.107 \\
                (c) & tetraaminooxalate & (C$_2$N$_4$H$_4$)$^{2-}$ & 0.215 & 1.526 & 5.091 \\
                (d) & tetrathiooxalate & (C$_2$S$_4$)$^{2-}$ & 0.238 & 1.201 & 3.012
        \end{tabular}
\end{ruledtabular}
\end{table}

\begin{figure}
\centering
\includegraphics[width=14cm]{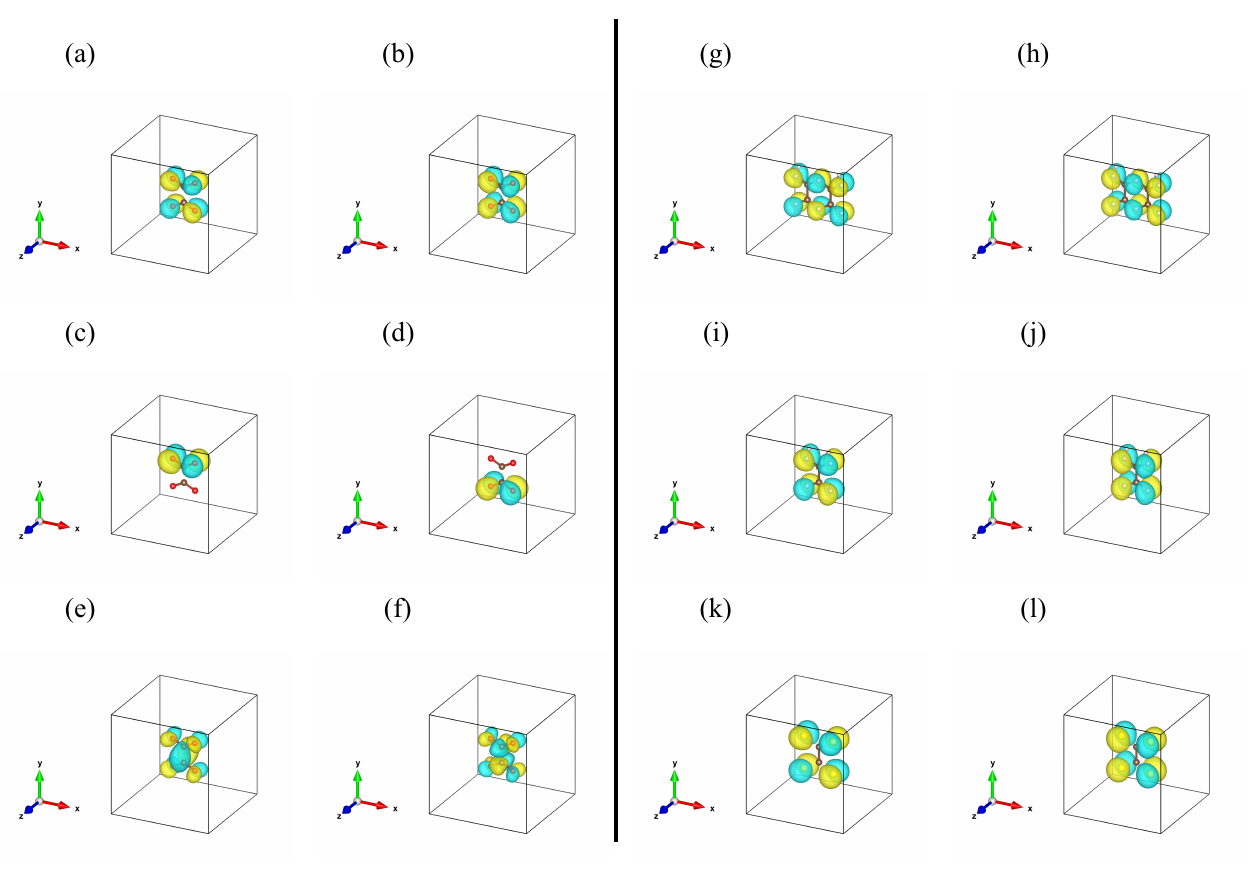}
\caption{Molecular orbitals of ligands. (a) Antisymmetric HOMO$-3$ of oxalate created by $a^\dagger=(u^\dagger-l^\dagger)/\sqrt{2}$ with an energy $E=-2.089$ eV. (b) Symmetric HOMO$-4$ of oxalate created by $b^\dagger=(u^\dagger+l^\dagger)/\sqrt{2}$ with $E=-2.396$ eV. (c) Upper-half fMO of oxalate created by $u^\dagger.$ (d) Lower-half fMO of oxalate created by $l^\dagger.$ (e) Symmetric LUMO of oxalate with $E=2.6$ eV (criterion). (f) Antisymmetric LUMO$+1$ of oxalate with $E=5.862$ eV. (g) Antisymmetric HOMO of (C$_4$N$_6$H$_4$)$^{2-}$ with $E=-1.418$ eV. (h) Symmetric HOMO$-1$ of (C$_4$N$_6$H$_4$)$^{2-}$ with $E=-1.833$ eV. (i) Antisymmetric HOMO of tetraaminooxalate with $E=-1.531$ eV. (j) Symmetric HOMO$-1$ of tetraaminooxalate with $E=-1.961$ eV. (k) Antisymmetric HOMO$-2$ of tetrathiooxalate with $E=-1.497$~eV. (l) Symmetric HOMO$-5$ of tetrathiooxalate with $E=-1.973$ eV. In each panel, yellow and blue bubbles represent plus and minus isosurfaces, respectively (C: brown, O: red, H: white, N: light blue, C: yellow for atoms).}
\label{mo}
\end{figure}

In the case of oxalate shown in Fig.~S\ref{mf}(a),
the $\pi$-conjugated highest occupied molecular orbitals (HOMOs) localized along the boundaries consist mostly of oxygen $p_z$-orbitals and are well-separated from each other.
We can thus easily express these orbitals by linear combinations of upper-half and lower-half ``fragment molecular orbitals'' (fMOs) localized
along each boundary.
(Here we regard the $y$-direction in Fig.~S\ref{mo} as the vertical direction,
viewing from the $z$-direction.)
The calculated HOMO, HOMO$-1$ and HOMO$-2$ are $\sigma$-orbitals with oxygen $s$-orbitals and thus
 ignored here, because they cannot hybridize with the Ru $t_{2g}$-orbitals.
(HOMO$-n$ refers to the molecular orbital $n$-th in the energy
below the highest occupied one.)
The calculated HOMO$-3$ shown in Fig.~S\ref{mo}(a)
(resp. HOMO$-4$ shown in Fig.~S\ref{mo}(b)) is an
antisymmetric (resp. symmetric) $\pi$-conjugated HOMO ($\pi$-HOMO in short) of oxalate and we define the
creation operator for this state as $a^\dagger$
(resp. $b^\dagger$). Actually, we could decompose this orbital by
$a^\dagger=(u^\dagger-l^\dagger)/\sqrt{2}$
(resp. $b^\dagger=(u^\dagger+l^\dagger)/\sqrt{2}$) using the upper-half
fMO $u^\dagger$ shown in Fig.~S\ref{mo}(c) and the
lower-half fMO $l^\dagger$ shown in Fig.~S\ref{mo}(d). If we
regard the potential energy of the two fMOs as $V_\pi$ and define a
hopping $-t_{\pi\pi}$ between these two orbitals, then it
can easily be concluded that HOMOs $a^\dagger (b^\dagger) = (u^\dagger
\mp l^\dagger)/\sqrt{2}$ have an energy of $V_\pi \pm t_{\pi\pi},$
respectively. From the calculated energy of HOMO$-3$ and HOMO$-4,$ we
can estimate the value of $V_\pi$ and $t_{\pi\pi}.$ Similarly, we can
decompose the $\pi$-conjugated lowest unoccupied molecular orbitals
(LUMOs) (LUMO and LUMO$+1$ shown in Fig.~S\ref{mo}(e) and
Fig.~S\ref{mo}(f), respectively, in the case of oxalate) into two pieces and estimate the
value of the potential energy $V_{\pi^*}$ and the hopping
$t_{\pi^*\pi^*}$ for these LUMOs. (LUMO$+n$ refers to the molecular orbital $n$-th in the energy
above the lowest unoccupied one.)  In Table~\ref{param},
we summarize the value of $t_{\pi\pi},$ $t_{\pi^*\pi^*},$ and
$V_{\pi^*}-V_\pi$ obtained in this way for each ligand molecule.  The
absolute values of the potential energy $V_{\pi}$ and $V_{\pi^*}$
obtained in the present calculation are not quite meaningful if molecules
are included in actual MOFs.  However, the difference $V_{\pi^*}-V_\pi$ is physical
and thus is listed in Table~\ref{param}.

In the case of (C$_4$N$_6$H$_4$)$^{2-}$ shown in Fig.~S\ref{mf}(b), the two $\pi$-HOMOs shown in Fig.~S\ref{mo}(g)
and Fig.~S\ref{mo}(h) are just below the Fermi energy even in the vacuum
and there is no irrelevant $\sigma$-HOMO between the Fermi energy and
these $\pi$-HOMOs.
In the case of tetraaminooxalate shown in Fig.~S\ref{mf}(c),
the two $\pi$-HOMOs shown in Fig.~S\ref{mo}(i) and Fig.~S\ref{mo}(j) have the same property.

Since tetrathiooxalate shown in Fig.~S\ref{mf}(d) might be able to coordinate almost octahedrally to the metal ion $M^{3+},$ ($M$ = Ru, Os, etc.)
i.e. \mbox{$\angle$ (S --- $M$ --- S) $\sim 90\degree$}, the replacement of $E$ = O
by S could possibly lead to a better ligand field~\cite{Itoi}
for the Jackeli-Khaliullin mechanism~\cite{Jackeli},
in terms of coupling to the $M$ = Ru $t_{2g}$-orbitals.
On the other hand,
the DFT calculation suggests that the $\pi$-HOMOs shown in
Fig.~S\ref{mo}(k) and Fig.~S\ref{mo}(l) are not as degenerate in
tetrathiooxalate as in oxalate.
This is rather disadvantageous for the cancellation of the Heisenberg
interaction due to the Jackeli-Khaliullin mechanism, as we discussed
in the main text.
By comparing Table~\ref{param} and Fig.~S\ref{mo}, we can see that $t_{\pi\pi}$
grows (O $<$ NH $<$ S) as the wavefunction around chalcogen (or NH)
expands.

\begin{figure}
\centering
\includegraphics[width=14cm]{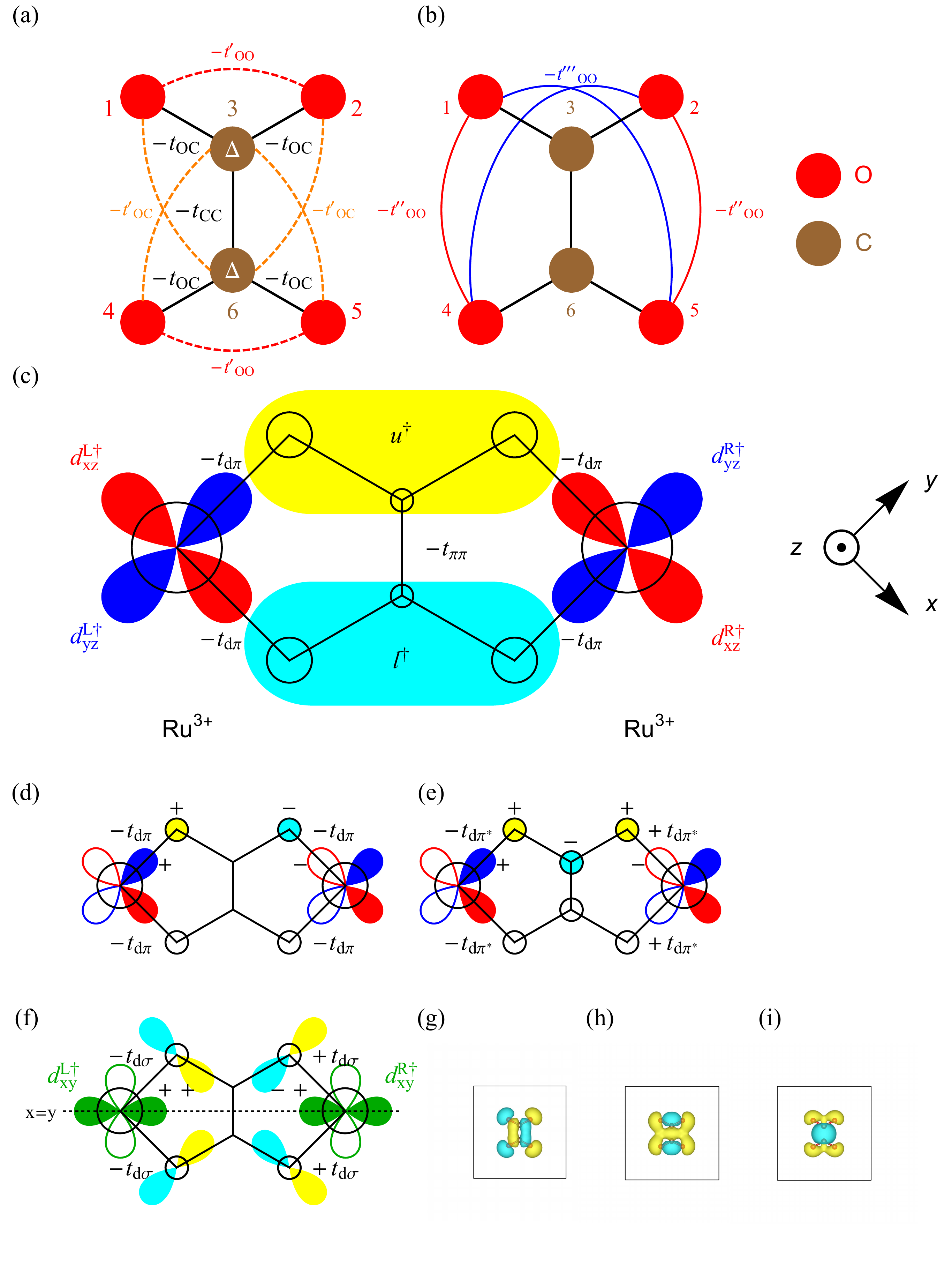}
\caption{Tight-binding model for superexchange interactions. (a) Tight-binding model for the oxalate ion with NN and 2NN interactions. (b) 3NN and 4NN interactions which break the degeneracy. (c) Tight-binding model for the superexchange pathways between the Ru $t_{2g}$-orbitals.
(d) Upper-half superexchange pathway for the $\pi$-HOMOs of oxalate. (e) Upper-half superexchange pathway for the $\pi$-LUMOs of oxalate. The signs of the contribution to the spin model are different between the two because the superposition coefficients of the oxygen $p_z$-orbitals are different in the DFT calculations (compare Fig.~S\ref{mo}(a) and Fig.~S\ref{mo}(b) to Fig.~S\ref{mo}(e) and Fig.~S\ref{mo}(f)).
(f) Possible superexchange pathway between the Ru $d_{xy}$-orbitals. (g) HOMO$-8$ (deep $\sigma$-orbital) of oxalate with $E=-6.150$ eV when viewed from the $z$-direction, which has the symmetry to hybridize to the Ru $d_{xy}$-orbitals.
(h) HOMO$-10$ with $E=-6.628$ eV. (i) HOMO$-12$ with $E=-10.40$ eV.}
\label{figs}
\end{figure}

\section*{S\lowercase{ection}~C: Edge states from the tight-binding model}

In order to understand the physical origin of the nearly-degenerate
$\pi$-HOMO states, which are important for the realization
of the Jackeli-Khaliullin mechanism,
we discuss a tight-binding model (or LCAO (linear combination of atomic orbitals) H\"uckel approximation) for the ligand molecules.
As an example, here we discuss the simplest case of oxalate
as the ligand molecule.
We consider the six 2$p_z$-orbitals of C and O atoms of oxalate.
Assuming the planar structure and $D_{2h}$ symmetry of oxalate, the tight-binding hopping Hamiltonian in the first-quantized form should be written in the following $6\times 6$ matrix.
\begin{equation}
H_\textrm{ox}=
\begin{pmatrix}
0 & -t_\textrm{OO}^\prime & -t_\textrm{OC} & -t_\textrm{OO}^{\prime\prime} & -t_\textrm{OO}^{\prime\prime\prime} & -t_\textrm{OC}^\prime \\
 -t_\textrm{OO}^\prime & 0 & -t_\textrm{OC} & -t_\textrm{OO}^{\prime\prime\prime} & -t_\textrm{OO}^{\prime\prime} & -t_\textrm{OC}^\prime \\
 -t_\textrm{OC} & -t_\textrm{OC} & \Delta & -t_\textrm{OC}^\prime & -t_\textrm{OC}^\prime & -t_\textrm{CC}\\
 -t_\textrm{OO}^{\prime\prime} & -t_\textrm{OO}^{\prime\prime\prime} & -t_\textrm{OC}^\prime & 0 & -t_\textrm{OO}^\prime & -t_\textrm{OC} \\
 -t_\textrm{OO}^{\prime\prime\prime} & -t_\textrm{OO}^{\prime\prime} & -t_\textrm{OC}^\prime & -t_\textrm{OO}^\prime & 0 & -t_\textrm{OC} \\
 -t_\textrm{OC}^\prime & -t_\textrm{OC}^\prime & -t_\textrm{CC} & -t_\textrm{OC} & -t_\textrm{OC} & \Delta
\end{pmatrix},
\end{equation}
where $\Delta$ is the difference in the potential energy between O and C, the 2$p_z$-orbital of each atom is numbered and the real-valued hopping parameters are defined as shown in Fig.~S\ref{figs}(a) and Fig.~S\ref{figs}(b).

First, we ignore the further-neighbor interactions by setting
$t_\textrm{OO}^\prime=t_\textrm{OO}^{\prime\prime}=t_\textrm{OO}^{\prime\prime\prime}=t_\textrm{OC}^\prime=0,$
and consider only the potential energy and the nearest-neighbor (NN)
interactions. In this case, by diagonalizing the Hamiltonian, we get one
pair of positive-energy modes, another pair of negative-energy modes,
and two degenerate zero-energy modes, $(1,-1,0,0,0,0)/\sqrt{2}$ and
$(0,0,0,1,-1,0)/\sqrt{2}$ as energy eigenstates.
In the present material, $\pi$-conjugated molecular orbitals are
$2/3$-filled, and thus these degenerate pairs of states correspond
to $\pi$-HOMOs.
Namely, the double degeneracy of the $\pi$-HOMO states is
exact in the limit where only NN hoppings are allowed.
These two degenerate
zero modes are similar to the $k=\pi$ zigzag edge modes of
graphene, and thereby localized on the O atoms.
Therefore, the (nearly) degenerate $\pi$-HOMOs can be physically
interpreted as ``edge states'' along the upper
and lower sides of the ligand molecule.

Let us now discuss the effects of further-neighbor hoppings.
We find that, because of the symmetry of the molecule,
the second-nearest-neighbor (2NN) interactions
$t_\textrm{OO}^\prime$ and $t_\textrm{OC}^\prime$ do not
lift the degeneracy of $\pi$-HOMOs even though they may shift
their overall energy.
Thus, the degeneracy is only lifted by introducing the
third-nearest-neighbor (3NN) hopping $t_\textrm{OO}^{\prime\prime}$
or the fourth-nearest-neighbor (4NN) hopping
$t_\textrm{OO}^{\prime\prime\prime}$ shown in Fig.~S\ref{figs}(b) with an energy split of
$2t_{\pi\pi}=2(t_\textrm{OO}^{\prime\prime}-t_\textrm{OO}^{\prime\prime\prime}).$
Therefore, in terms of the tight-binding model,
we can identify the effective tunneling
matrix element $t_{\pi\pi}$ between the two $\pi$-HOMOs
with
$t_\textrm{OO}^{\prime\prime}-t_\textrm{OO}^{\prime\prime\prime}.$
It is naturally expected that this value would be smaller than the NN
and 2NN interactions, as they involve longer-range hoppings.
From the DFT calculation, we obtain
$t_{\pi\pi}=0.153$ eV.
The smallness of this value, compared to the other energy scale
$V_{\pi^*}-V_\pi = 6.474$ eV, indeed reflects this mechanism.

Since the 4NN hopping $t_\textrm{OO}^{\prime\prime\prime}$
would be suppressed more strongly than the
3NN one $t_\textrm{OO}^{\prime\prime},$
we can approximately identify
$t_\textrm{OO}^{\prime\prime} \sim t_{\pi\pi}=0.153$ eV.
We note that, even this small
value of $t_{\pi\pi}$ may be an overestimation in our calculation
done for the ligand molecule in vacuum; 
in the actual metal-organic frameworks, there is an electron
density from the Ru atoms between the upper and lower oxygens, which would
screen the hopping between the oxygens and thus
suppress $t_\textrm{OO}^{\prime\prime}.$
The analysis here applies to all of the oxalate family;
the nearly-degenerate $\pi$-HOMOs can be similarly understood as
``edge states,'' which are exactly degenerate within the tight-binding
approximation with the NN and 2NN hoppings.

Finally, we comment on the tetraaminopyrazine case in the
qualitative level. In the analysis similar to the oxalate case,
we find two edge states also for this ligand molecule.
However, unlike the oxalate case,
these states are split due to the incomplete
degeneracy of the renormalized
potential energy of the three N atoms, or the discrepancy between the
2NN interactions between the N and C atoms.
As a consequence, $t_{\pi\pi}$ should be larger in
the tetraaminopyrazine case than in the oxalate case, which agrees with
the DFT results shown in Table~\ref{param}. We also note that there is no
reason to expect near-degeneracy in dhbq$^{2-}$ or $X_2$An$^{2-}$ (dhbq = 2,5-dihydroxy-1,4-benzoquinone, $X$ = Cl, Br, etc. and An = anilate) and the
difference between these ligands and the tetraaminopyrazine-based ligand
comes from the substantial potential difference between the chalcogen $p_z$-orbital
and the C $p_z$-orbital.

\section*{S\lowercase{ection}~D: Derivation of the JK$\Gamma$ model}

We construct a microscopic
model for the superexchange between the Ru $t_{2g}$-orbitals via the planar organic ligand
in the $xy$ plane using the fMO method.  We mostly consider superexchange pathways through
the $\pi$-conjugated system, i.e. $p_z$-orbitals. Actually, the symmetry allows hybridization
between the Ru $t_{2g}$-orbitals and the $p_x$ or $p_y$-orbitals of chalcogens (or nitrogen)
and they can form $\sigma$-HOMOs or $\sigma^*$-LUMOs of the organic ligand. However,
the DFT results suggest that such orbitals will not form any superexchange pathways near
the Fermi energy for oxalate and tetraaminopyrazine, and there only remains
$\sigma$-HOMOs or $\sigma^*$-LUMOs with $s$-orbitals (or the orbitals with the same
symmetry when viewed from Ru) on oxygens (in the case of oxalate), which cannot be
hybridized with the Ru $t_{2g}$-orbitals near the Fermi level. For oxalate, the first
$\sigma$-HOMO that contributes to the superexchange interaction is HOMO$-8$ far below
the Fermi energy and we mostly consider the contribution of this orbital from $\sigma$-HOMOs.

For $\pi$-HOMOs, we divide the $\pi$-conjugated molecular orbital of the organic ligand
into the upper-half fMO with a creation operator $u^\dagger$ and the lower-half fMO with
a creation operator $l^\dagger,$ as shown in Fig.~S\ref{figs}(c). If we define a real-valued
hopping amplitude $-t_{\pi\pi}$ between these two fMOs, it can easily be concluded that
HOMOs $(u^\dagger \pm l^\dagger)/\sqrt{2}$ would be split with the energy of
$V_{\pi} \mp t_{\pi\pi},$ respectively, assuming the potential energy of the $\pi$-HOMOs
as $V_{\pi}.$ If we regard $u^\dagger$ and $l^\dagger$ as $p_z$-orbitals of two oxygens
between Ir$^{4+}$ in iridates~\cite{Jackeli}, we can conclude that the superexchange
interaction would be completely Kitaev-type by the Jackeli-Khaliullin mechanism when
$t_{\pi\pi}=0.$ It is known that the Kitaev interaction comes from the off-diagonal
hopping between $t_{2g}$-orbitals, such as between $d_{xz}$ and $d_{yz},$ while the
Heisenberg interaction mainly comes from the diagonal part. The diagonal element of
the hopping matrix always needs the hopping between $u^\dagger$ and $l^\dagger$
somewhere in the superexchange pathway, so it must be important to newly include
the effect of $t_{\pi\pi}$ to estimate the interactions other than the Kitaev term.
By comparing with the results in the previous section, we can regard the
$u^\dagger$ state as $(1,-1,0,0,0,0)/\sqrt{2}$ and the $l^\dagger$ state
as $(0,0,0,1,-1,0)/\sqrt{2}$ in oxalate.
We note that the present fMO analysis does not require the near-degeneracy
of $\pi$-HOMOs, although the near-degeneracy is advantageous for
the Jackeli-Khaliullin mechanism.
In fact, in the explicit derivation of the effective model,
we can see how the near-degeneracy is important for the suppression
of the Heisenberg term.

The model Hamiltonian in the second-quantized form is
\begin{equation}
        H = -t_{d\pi} (u^\dagger d_{yz}^L+l^\dagger d_{xz}^L+u^\dagger d_{xz}^R+l^\dagger d_{yz}^R+h.c.) +V_\pi(u^\dagger u+l^\dagger l)-t_{\pi\pi} (u^\dagger l+l^\dagger u),
\end{equation}
where $-t_{d\pi}$ is a real-valued hopping element between the Ru $t_{2g}$-orbitals and
the fMO, $V_\pi$ is a potential energy which electrons from Ru feel on the fMOs, and
$d_{i}^L$ (resp. $d_{i}^R$) is the annihilation operator of an electron on the
Ru $d_i$-orbital on the left (resp. right) side in Fig.~S\ref{figs}(c).

We define $b^\dagger=(u^\dagger+l^\dagger)/\sqrt{2}$ and $a^\dagger=(u^\dagger-l^\dagger)/\sqrt{2}$ to diagonalize the $t_{\pi\pi}$ term,
\begin{equation}
        H = -\frac{t_{d\pi}}{\sqrt{2}}[(a^\dagger+b^\dagger)d_{yz}^L+(b^\dagger-a^\dagger)d_{xz}^L+(a^\dagger+b^\dagger)d_{xz}^R+(b^\dagger-a^\dagger)d_{yz}^R+h.c.]+(V_\pi+t_{\pi\pi})a^\dagger a+(V_\pi-t_{\pi\pi})b^\dagger b.
\end{equation}
Then, we get an effective hopping matrix via $\pi$-HOMOs $H_\textrm{dd}^\pi$ between
the Ru $d$-orbitals from the second-order perturbation in $t_{d\pi}/|V_\pi\pm t_{\pi\pi}|$
by integrating out the $b$ and $a$ states.
\begin{equation}
        H_{dd}^{\pi} = -\frac{t_{d\pi}^2 t_{\pi\pi}}{V_\pi^2-t_{\pi\pi}^2} (d_{yz}^{L\dagger} d_{yz}^R+d_{xz}^{L\dagger} d_{xz}^R)-\frac{t_{d\pi}^2 V_\pi}{V_\pi^2-t_{\pi\pi}^2} (d_{yz}^{L\dagger} d_{xz}^R+d_{xz}^{L\dagger} d_{yz}^R) +h.c.
\end{equation}

We can repeat a similar calculation to derive the hopping matrix via the $\pi^*$-LUMOs
of the organic ligand. If we define a potential energy $V_{\pi^*}$ for the LUMOs,
a real-valued hopping $-t_{\pi^*\pi^*}$ between the LUMOs and a real-valued hopping
$-t_{d\pi^*}$ between the Ru $t_{2g}$-orbitals and the LUMO, we get an effective
hopping matrix $H_{dd}^{\pi^*}$ as
\begin{equation}
        H_{dd}^{\pi^*} = \frac{t_{d\pi^*}^2 t_{\pi^*\pi^*}}{V_{\pi^*}^2-t_{\pi^*\pi^*}^2} (d_{yz}^{L\dagger} d_{yz}^R+d_{xz}^{L\dagger} d_{xz}^R)+\frac{t_{d\pi^*}^2 V_{\pi^*}}{V_{\pi^*}^2-t_{\pi^*\pi^*}^2} (d_{yz}^{L\dagger} d_{xz}^R+d_{xz}^{L\dagger} d_{yz}^R) +h.c.
\end{equation}
We have to note that the HOMOs and the LUMOs contribute to the effective hopping with
different signs due to the difference in the superposition coefficients of the oxygen
$p_z$-orbitals (see Fig.~S\ref{figs}(d) for the HOMOs and Fig.~S\ref{figs}(e) for the LUMOs).
In addition, we consider $\sigma$-HOMO (HOMO$-8$ in oxalate) with a potential energy
$V_\sigma.$ The symmetry only allows the hopping $-t_{d\sigma}$ between the $\sigma$-HOMO
and the Ru $d_{xy}$-orbital as shown in Fig.~S\ref{figs}(f), and HOMO$-8$ shown
in Fig.~S\ref{figs}(g) actually possesses this property. We have to note that this energy
eigenstate orbital is symmetric against $x=y,$ and anti-symmetric HOMO$-7$ does not
contribute to the superexchange between the Ru $d_{xy}$-orbitals.
\begin{equation}
        H_{dd}^\sigma = \frac{(2t_{d\sigma})^2}{V_\sigma} (d_{xy}^{L\dagger} d_{xy}^R)+h.c.
\end{equation}
The coefficient $2$ comes from the symmetry against $x=y$ of the $\sigma$-HOMO.
Combining these three contributions and following Ref.~\cite{JKG2}, we estimate an effective hopping matrix
$H_\textrm{dd}^\textrm{eff}$ between the Ru $d$-orbitals~\footnote{In Ref.~\cite{JKG2}, there is another term $t_4$ from $xy$ to $xz$ or $yz,$ but we can ignore this contribution assuming the completely planar structure of molecule and the local mirror symmetry against this bond plane.  This symmetry defines the distinction between $\pi$- and $\sigma$-orbitals and $t_4$ vanishes when they are independent.}.
\begin{align}
        H_{dd}^\textrm{eff} &= H_{dd}^{\pi}+H_{dd}^{\pi^*}+H_{dd}^{\sigma} = -t_1 (d_{yz}^{L\dagger} d_{yz}^R+d_{xz}^{L\dagger} d_{xz}^R)-t_2 (d_{yz}^{L\dagger} d_{xz}^R+d_{xz}^{L\dagger} d_{yz}^R)-t_3 d_{xy}^{L\dagger} d_{xy}^R +h.c., \\
        t_1 &= \frac{t_{d\pi}^2 t_{\pi\pi}}{V_\pi^2-t_{\pi\pi}^2}-\frac{t_{d\pi^*}^2 t_{\pi^*\pi^*}}{V_{\pi^*}^2-t_{\pi^*\pi^*}^2}, \\
t_2 &= \frac{t_{d\pi}^2 V_\pi}{V_\pi^2-t_{\pi\pi}^2}-\frac{t_{d\pi^*}^2 V_{\pi^*}}{V_{\pi^*}^2-t_{\pi^*\pi^*}^2}, \\
t_3 &= -\frac{4t_{d\sigma}^2}{V_\sigma}+\dots
\end{align}
Of course, $V_\pi<0$ and $V_{\pi^*}>0$ (while $t_{\pi\pi}>0$ and $t_{\pi^*\pi^*}>0$),
so the $\pi$-HOMOs and the $\pi^*$-LUMOs contribute to the diagonal hopping $t_1$ with
the opposite sign and the off-diagonal hopping $t_2$ with the same sign, which makes
$|t_1|/|t_2|$ even smaller.

Therefore, by mapping the electron hopping matrix to the $J_\textrm{eff}=1/2$ spin model
by the second-order perturbation in $t_i,$ we actually get the JK$\Gamma$ model for the
bond in the $xy$-plane, which is defined in Eq.~(3) in the main text. A similar analysis
is possible for bonds in the other directions and we can construct a whole JK$\Gamma$
model for both two-dimensional (2D) and three-dimensional (3D) lattices.
We note that in chiral 3D tricoordinated lattices
the parity symmetry is explicitly broken and we cannot ignore Dzyaloshinskii-Moriya interactions.
Assuming zero crystal
field splitting for the Ru $t_{2g}$-orbitals and ideal octahedral coordination,
the parameters for the JK$\Gamma$ model will be the following~\cite{JKG2}.
\begin{align}
        J &= \frac{4\mathbb{A}}{9}(2t_1+t_3)^2-\frac{16\mathbb{B}}{9} (t_1-t_3)^2, \\
        K &= \frac{8\mathbb{B}}{3} [(t_1-t_3)^2-3t_2^2], \\
        \Gamma &= \frac{16\mathbb{B}}{3} t_2(t_1-t_3),
\end{align}
where the parameters $\mathbb{A}$ and $\mathbb{B}$ can be expressed in terms with
a Hund coupling $J_H,$ a spin-orbit coupling $\lambda,$ and a Hubbard term $U$ for
Ru$^{3+}$ in \cite{JKG2} as follows.
\begin{align}
        \mathbb{A} &= -\frac{1}{3} \Bigl[ \frac{J_H+3(U+3\lambda)}{6J_H^2-U(U+3\lambda)+J_H(U+4\lambda)} \Bigr], \\
        \mathbb{B} &= \frac{4}{3} \Bigl[ \frac{(3J_H-U-3\lambda)}{(6J_H-2U-3\lambda)}\eta\Bigr], \\
        \eta &= \frac{J_H}{6J_H^2-J_H(8U+17\lambda)+(2U+3\lambda)(U+3\lambda)}.
\end{align}

Then, the literature~\cite{JKG2} estimates these parameters~\footnote{Ref.~\cite{JKG2} defines these values for 4$d^5$ metal ions in general.  Therefore, we can expect that these values can be used not only for $\alpha$-RuCl$_3,$ but also for MOFs.} for Ru$^{3+}$ as
$\mathbb{A}\sim 0.6$ eV$^{-1}$ and $\mathbb{B}\sim 0.05$ eV$^{-1}.$
If we only consider the superexchange pathways through $\pi$-HOMOs (i.e. $t_3 \sim 0$), from the relations
$t_1/t_2=t_{\pi\pi}/V_\pi$ and $\mathbb{B}/\mathbb{A}\sim 1/10,$ we can conclude that,
if $|t_{\pi\pi}|/|V_\pi| \sim 1/10$ for two almost degenerate HOMOs, the superexchange
interaction via these HOMOs should be Kitaev-dominant,
i.e. $J/|K| \sim |\Gamma|/|K| \sim 1/10.$
Thus, we have shown that $|t_{\pi\pi}|/|V_\pi|$ actually controls the parameters of
the JK$\Gamma$ model as expected. Next, if we consider the contribution from
the $\pi^*$-LUMOs, then
\begin{equation}
        \frac{t_{d\pi}^2 t_{\pi\pi}}{V_\pi^2} \gtrsim \frac{t_{d\pi^*}^2 t_{\pi^*\pi^*}}{V_{\pi^*}^2}
\end{equation}
must be held in order not to change the order of $|t_1|/|t_2|.$
We can roughly estimate $t_{d\pi^*}/t_{d\pi}$ as follows.
If we assume only the Ru 4$d_{yz}$-orbital $\ket{\psi_{\textrm{4}d_{yz}}}$ and
the neighboring O 2$p_z$-orbital $\ket{\psi_{\textrm{2}p_z}}$ contribute to the
hopping element, using the real-valued LCAO coefficient $c_\pi$ (resp. $c_{\pi^*}$)
of the the neighboring O 2$p_z$-orbital for the HOMO (resp. LUMO), we can define a
hopping element $-t_{dp}=\bra{\psi_{\textrm{2}p_z}} H_\textrm{LCAO} \ket{\psi_{\textrm{4}d_{yz}}},$ where $H_\textrm{LCAO}$ is a hopping Hamiltonian in the LCAO approximation. Then,
\begin{equation}
        \frac{-t_{d\pi^*}}{-t_{d\pi}} = \frac{-c_{\pi^*}t_{dp}}{-c_\pi t_{dp}} = \frac{c_{\pi^*}}{c_\pi}.
\end{equation}
Using this relation, we can
estimate the $t_{d\pi^*}/t_{d\pi}$ for oxalate from the LCAO coefficients calculated
by \textsc{openmx}, assuming the almost orthogonality of the pseudoatomic orbitals, as
\begin{equation}
        \frac{t_{d\pi^*}}{t_{d\pi}}=\frac{c_{\pi^*}}{c_\pi}=\frac{\sqrt{2}\cdot 0.3063}{\sqrt{2}\cdot 0.4972}=0.6159.
\end{equation}

\begin{figure}[H]
\centering
\includegraphics[width=10cm]{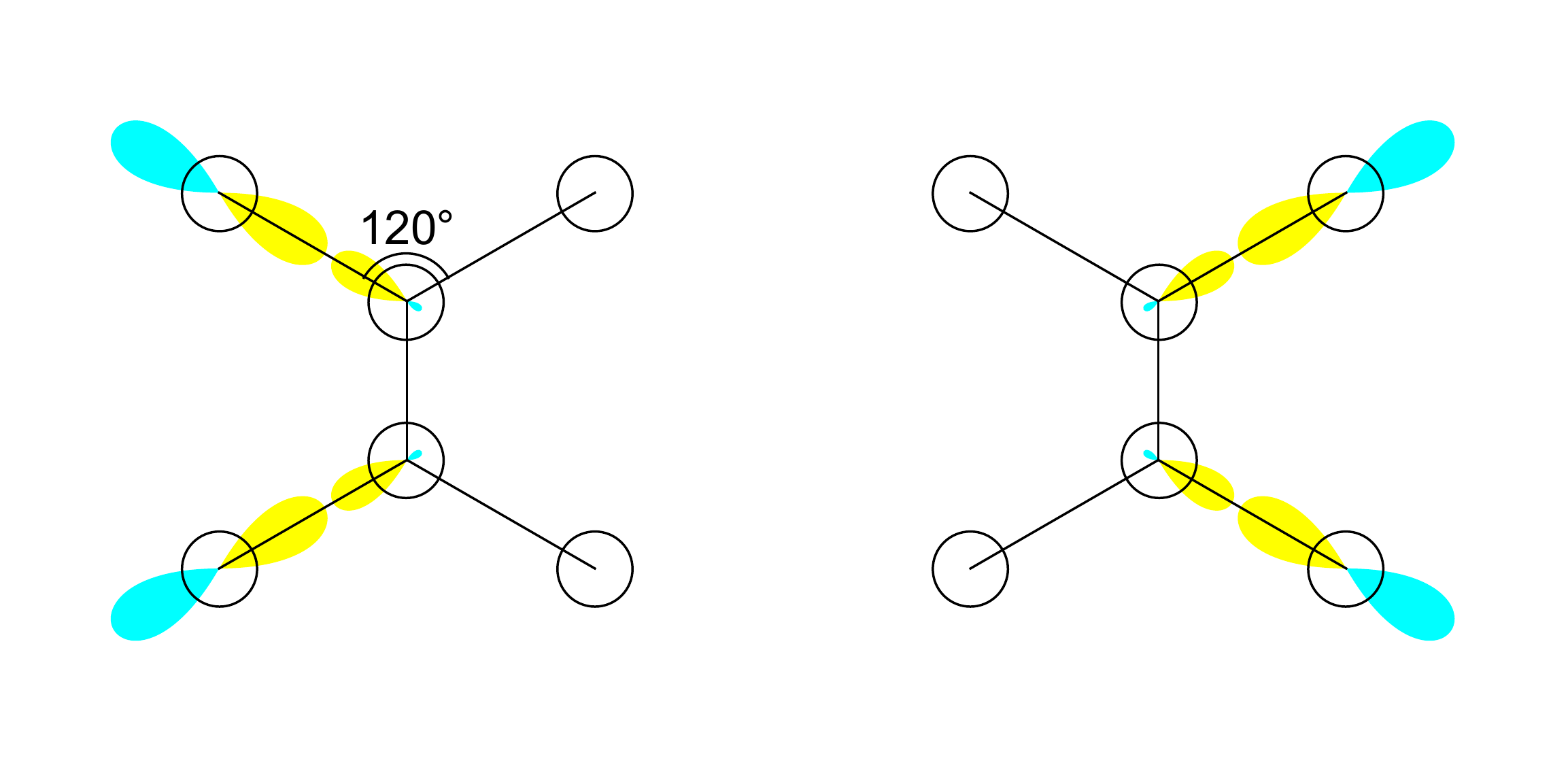} 
\caption{Two almost degenerate $\sigma$-orbitals localized along the left and right edges. They consist of the O-$p_x$ or $p_y$ orbitals and the C-$sp^2$ hybridized orbitals.}
\label{sp2}
\end{figure}

Finally, we estimate the possible contribution to $t_3$ from the localized $\sigma$-orbitals.
This contribution to $t_3$ should be smaller than the dominant contribution from
delocalized $\pi$-orbitals to $t_2$ because the $\sigma$-orbitals form covalent bonds
and the paired electrons are localized on those bonds, i.e. a charge gap is opened
for these localized valence bonds.  However, there is still a possibility that
the magnitude of $t_3$ is comparable to that of $t_1,$ so we made a crude estimation
for this contribution. As we explained above, $\sigma$-HOMOs do not contribute to
the hopping near the Fermi level, so, as the first approximation, we only consider
the effects of the deep levels, HOMO$-8$, HOMO$-10$, and HOMO$-12$ shown
in Fig.~S\ref{figs}(g), (h) and (i), respectively, with $|V_\sigma|>|V_\pi|.$
These orbitals can contribute only to $t_3$ due to the completely planar structure
of the molecule,
and their contributions with different signs
almost cancel out. This fact can be understood in a way similar to the
earlier analysis of the degenerate $\pi$-HOMOs.
From the viewpoint of localized valence bonds, two $\sigma$-orbitals localized
along the left and right edge shown in Fig.~S\ref{sp2} are almost orthogonal and
degenerate assuming the 120 degrees bonding around C and the complete degeneracy
of the C-$sp^2$ hybridized orbitals. If these orbitals are completely degenerate
and localized along the edges, they do not provide any superexchange pathways.
In the language of molecular orbitals, the contributions from $\sigma$-HOMOs created
by the linear combination of these two edge states will be cancelled out.
In the same way, we can naturally expect that the superexchange contributions
from nearly degenerate HOMO$-8,$ HOMO$-10,$ and HOMO$-12$ are rather destructive,
even if the degeneracy of the edge state is not perfect.
The relative values of $t_{d\sigma n}$ for HOMO$-n$ can be estimated in the same
way as $t_{d\pi}$ as follows.
\begin{align}
        \frac{t_3}{t_2} &=\Bigl(-\frac{4t_{d\sigma 8}^2}{V_{\sigma 8}}+\frac{4t_{d\sigma 10}^2}{V_{\sigma 10}} +\frac{4t_{d\sigma 12}^2}{V_{\sigma 12}}\Bigr)/t_2 \sim -0.196, \\
        \frac{t_{d\sigma 8}}{t_{d\pi}} &=\frac{0.4018}{\sqrt{2}\cdot 0.4972} = 0.5714,\quad
        \frac{t_{d\sigma 10}}{t_{d\pi}} =\frac{0.2398}{\sqrt{2}\cdot 0.4972} = 0.3410,\quad
        \frac{t_{d\sigma 12}}{t_{d\pi}} =\frac{0.2149}{\sqrt{2}\cdot 0.4972} = 0.3056.
\end{align}

From these values we can estimate $t_3/t_2 \sim -0.196$ as written in the main text.
Therefore, we can conclude for oxalate that if
$|t_{\pi\pi}|/|V_\pi| \sim 1/10,$ $|V_{\pi}|/|V_{\pi^*}|\sim 1/2$ and the destructive
interference in the $\sigma$-orbitals works, the interaction should be Kitaev-dominant
even if we consider contributions other than the $\pi$-HOMOs.
Then, the ideal situation for the Kitaev-dominant interaction can
be depicted schematically as follows (ox$^{2-}$ is oxalate).
\begin{tcolorbox}[colback=white,title=Ideal situation for oxalate-based frameworks]
\centering
\begin{MOdiagram}[names,labels-fs=\footnotesize]
        \atom[Ru$^{3+}$]{left}{1s={;up}}
 \atom[Ru$^{3+}$]{right}{1s={;up}} \molecule[ox$^{2-}$]{
 1sMO={1.68/3.01;pair}, label = {1sigma* = {$\pi^*$}} }
 \AO(3cm){s}[label={$\pi$}]{-1.987;pair}

  \AO(3cm){s}{-0.3805;pair} \AO(3cm){s}{-0.6331;pair}
  \AO(3cm){s}{-1.471;pair} \AO(3cm){s}[label={$\pi^*$}]{6.271;empty}
  \node[right,xshift=2mm] at (AO3) {irrelevant $\sigma$};
  \draw[<->,semithick] (AO2.0) -- (AO4.0);
\end{MOdiagram}
\end{tcolorbox}

\begin{tcolorbox}[colback=white,title=Ideal situation for tetraaminopyrazine-based frameworks]
        \centering
\begin{MOdiagram}[names,labels-fs=\footnotesize]
        \atom[Ru$^{3+}$]{left}{1s={;up}}
 \atom[Ru$^{3+}$]{right}{1s={;up}} \molecule[(C$_4$N$_6$H$_4$)$^{2-}$]{
 1sMO={1.013/1.385;pair}, label = {1sigma* = {$\pi^*$}} }
 \AO(3cm){s}[label={$\pi$}]{-1.428;pair}
\end{MOdiagram}
\end{tcolorbox}
To achieve every condition, the energy level of the Ru
$t_{2g}$-orbitals should sit in the right position between the HOMOs and
the LUMOs of the proposed oxalate-based or tetraaminopyrazine-based
ligands.
It will be interesting to carry out more accurate analysis of
these materials to see whether the dominance of the Kitaev interaction is maintained.

\section*{S\lowercase{ection}~E: Bond anisotropy}

\begin{figure}
\centering
\includegraphics[width=7cm]{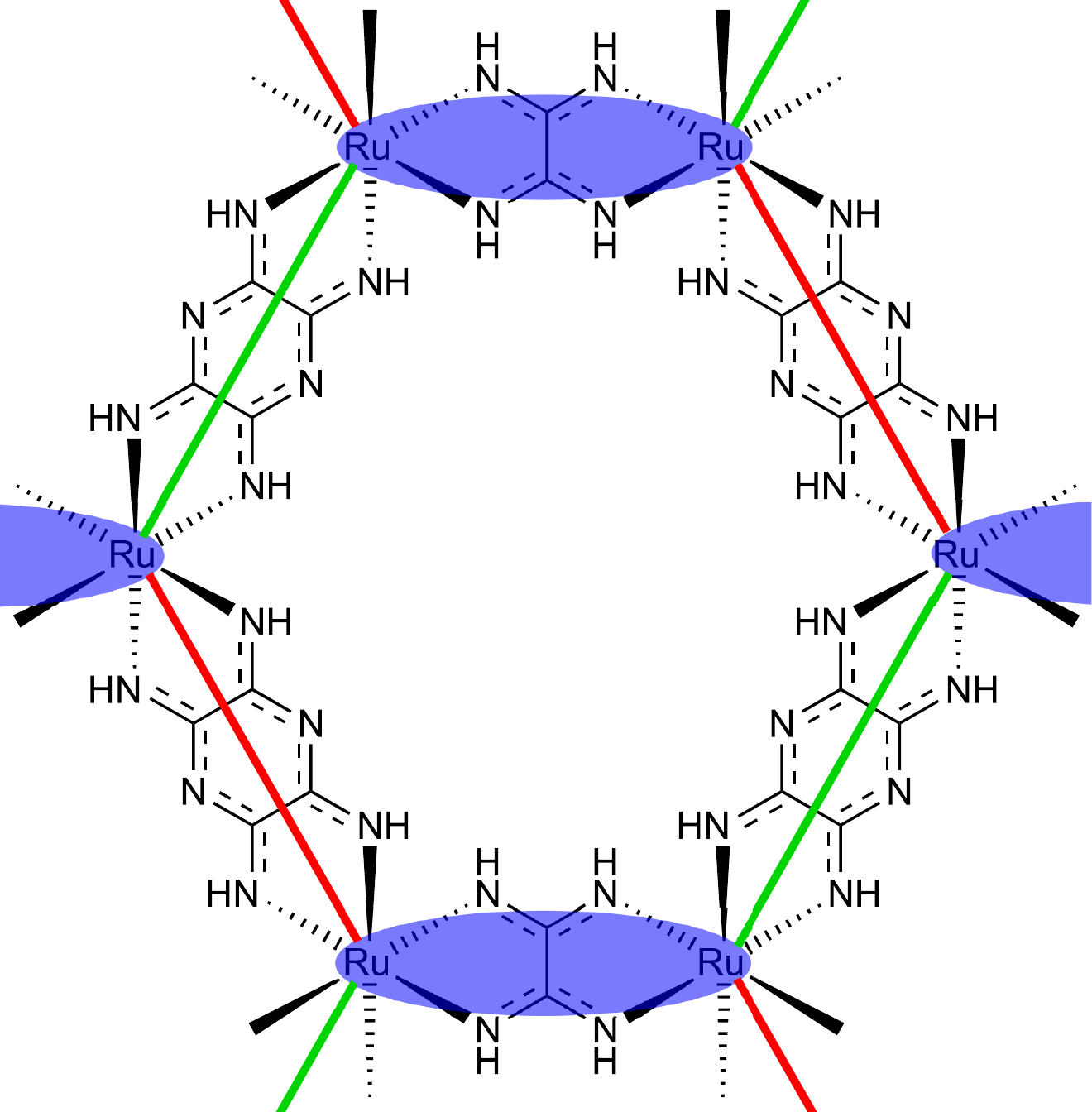} 
\caption{Heterogeneous distorted honeycomb structure to realize a gapped spin liquid region. The bond interaction is shown with the same color as in Fig.~1(a) in the main text and the Ru ions connected by a blue circle will be dimerized.}
\label{toric}
\end{figure}

Here we discuss MOFs with heterogeneous organic ligands, as
proposed in the section \textit{Designing
a Variety of Kitaev Spin Liquids} in the main text.
The heterogeneous organic ligands could arrange themselves
into an MOF with different ligand molecules depending on
the orientation of the bond, as shown in Fig.~S\ref{toric}.
This will naturally lead to the effective spin Hamiltonian with
bond anisotropy, namely the interaction strength depending on
the bond orientation.
In the following discussion, for simplicity, we
ignore the $\Gamma$ term of the JK$\Gamma$ model and assume only the
Kitaev-Heisenberg interaction with a bond anisotropy.
The effective spin Hamiltonian on the honeycomb lattice is then given by
\begin{equation}
        H=\sum_{\langle ij\rangle \in \alpha \beta(\gamma)} [J_\gamma \bm{S}_i \cdot\bm{S}_j-|K_\gamma| S_i^\gamma S_j^\gamma ],
\end{equation}
where $K_\gamma<0$ is the ferromagnetic Kitaev term and $J_\gamma$ is
the antiferromagnetic Heisenberg term for each type of bonds as shown in
Fig.~S\ref{toric}.
$\alpha,\beta,\gamma \in \{x,y,z\}$
and $\langle ij\rangle \in \alpha \beta(\gamma)$ 
mean that the bond plane of the nearest-neighbor bond $\langle ij\rangle$
is the $\alpha\beta$-plane perpendicular to the $\gamma$-axis.
We can expect $|K_z|\gg |K_x|=|K_y|
\gtrsim J_x=J_y$ and $|K_z|\gtrsim J_z$ because the bond length in the
$xy$-plane is shorter than the others. Then, we can treat the $K_z$ and
$J_z$ terms as unperturbed parts, and $K_x,$ $K_y,$ $J_x$ and $J_y$
terms as perturbation. In the zeroth-order ground state, if
$|K_z|>2J_z,$ the Ru ions connected by a blue circle in Fig.~S\ref{toric} will be dimerized ferromagnetically to
$\ket{\uparrow}\otimes\ket{\uparrow}$ or
$\ket{\downarrow}\otimes\ket{\downarrow}.$ The perfect Kitaev model with
$J_\gamma=0$ will be mapped to the 2D toric code~\cite{toriccode} if we
see these dimers as pseudospins up and down, respectively, by using
the fourth-order perturbation in $K_x/K_z$ and $K_y/K_z,$ and applying
local unitary transformations. The ground state of the 2D toric code has
a 2D $Z_2$ topological order and has fractionalized anyonic
excitations. Even in the 3D case, a similar distorted heterogeneous
structure is possible and the perfect 3D Kitaev model could be mapped to
the 3D toric code in the same way as in the 2D case~\cite{Mandal2}. This
model shows a 3D $Z_2$ topological order, which has string excitations
with exotic statistics.

If we include the Heisenberg term, the situation is known to be different~\cite{distorted}.
Even in the first order, there appears an antiferromagnetic Ising interaction
$J_xS_i^z S_j^z$ between the pseudospins of two nearest-neighbor dimers $i$ and $j.$
Therefore, in order for the toric code stabilizer term to be dominant, we first have to
impose the condition $K_x^2 K_y^2/(16|K_z|^3) \gg J_x$ in this toric-code limit
because the Ising term strongly favors a conventional N\'eel order for pseudospins,
i.e. a stripy order for original $J_{\textrm{eff}}=1/2$ spins.  In addition, there
appears a pseudospin flip interaction both from the second-order perturbation due
to the $J_z$ term and from the ignored $\Gamma$ term, so we have to investigate
whether or not these quantum fluctuations destroy the topological order similarly
to the thermal fluctuation~\cite{NGT2,NGT3} as an interesting future problem.
In the heterogeneous case, the
distortion due to the lack of the octahedral symmetry around Ru could make
the effective model deviate from the JK$\Gamma$ model more strongly than
in the homogeneous case, but we can still expect an almost octahedral
coordination if we use an amino-group in every direction as shown in
Fig.~S\ref{toric}.

\section*{S\lowercase{ection}~F: Lattice distortions in three-dimensional structures}

\begin{figure}
\centering
\includegraphics[width=8cm]{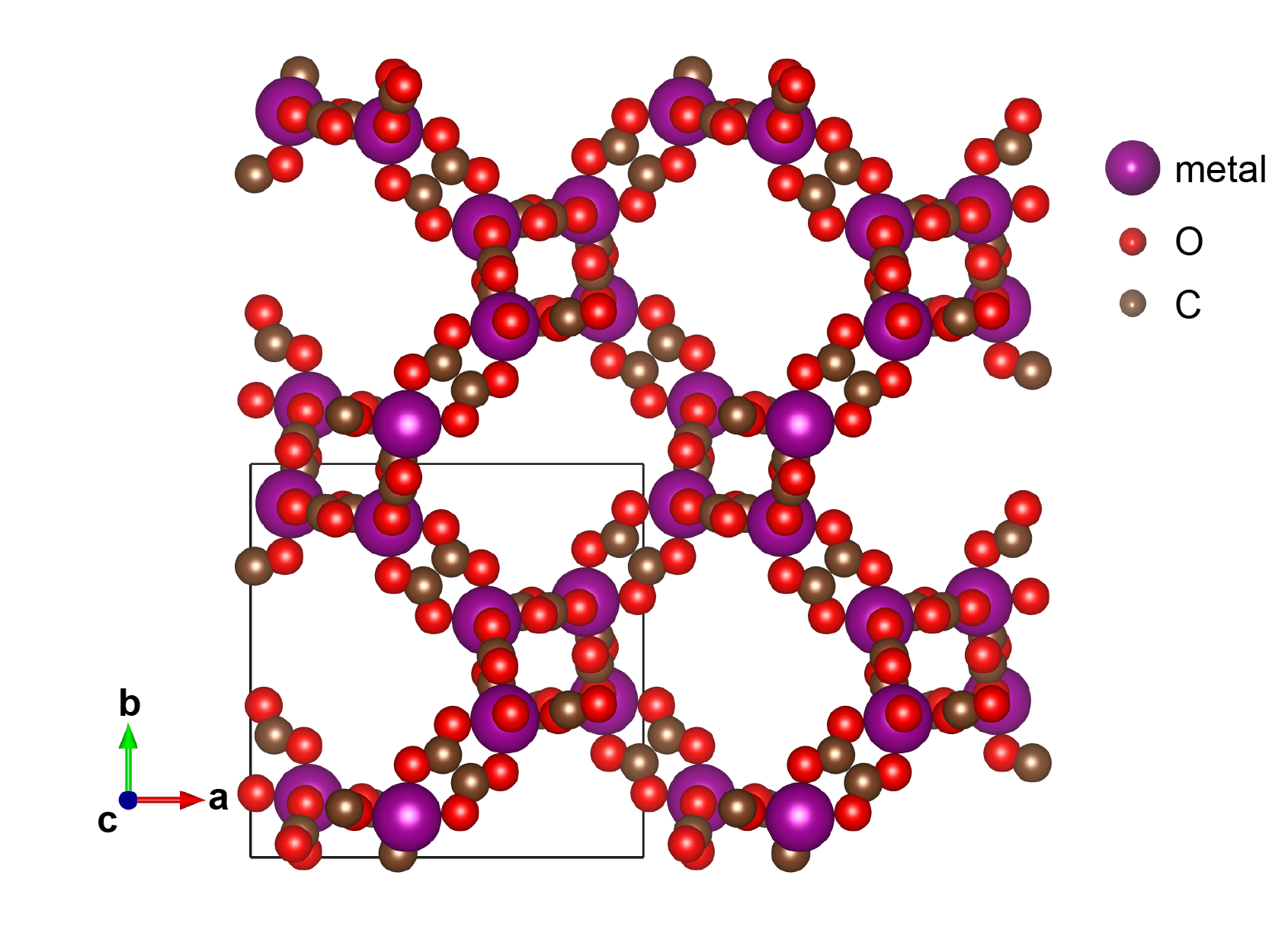} 
\caption{(10,3)-$a$ MOF with a $P4_1 32$ space group symmetry viewed from the
[001] direction.  While the ideal (10,3)-$a$ should look like the square-octagon lattice when
projected onto the (001) plane, the square spirals are tilted and, thus, regular
octagons are distorted in two ways alternately in this figure.  The unit cell is shown in the
black solid line and it is clear to see this distortion breaks the translational symmetry with
a vector $(\frac{1}{2},\frac{1}{2},\frac{1}{2}).$
This figure is reconstructed from the crystal data included in
Supporting Information of Ref.~\cite{Coronado2} (metal ions: purple, C: brown, O: red).}
\label{hyperoctagon}
\end{figure}

While a gapless nodal-line or Weyl spin liquid on (10,3)-$b$ (the 3D hyperhoneycomb lattice)
shown in Fig.~S\ref{hyper} is stable against small lattice distortions,
Majorana Fermi surfaces on (10,3)-$a$ (the 3D hyperoctagon lattice) need protection
by its $I4_1 32$ space group symmetry.  For example, a (10,3)-$a$ MOF with a space group
$P4_1 32,$ which is a subgroup of $I4_1 32,$ has been realized~\cite{Coronado2},
but the crystal system is lowered from a body-centered cubic system to a simple cubic system,
which means that the unit cell of this real material is doubled from the ideal
(10,3)-$a$ lattice, as shown in Fig.~S\ref{hyperoctagon}.
In this case, Fermi surfaces are unstable against an infinitesimal
perturbation away from the $K_x=K_y$ line, i.e. the Majorana Fermi surfaces
on (10,3)-$a$ have to be protected by the translational symmetry with a vector
$(\frac{1}{2},\frac{1}{2},\frac{1}{2})$~\cite{Maria2014}.
This difference between $I4_1 32$ and $P4_1 32$ comes from the fact that
the reduced unit cell of the $I4_1 32$ structure is incommensurate
with the bipartite coloring of (10,3)-$a.$
We still need a further materials search for an ideal $I4_1 32$ structure
to see quantum interaction effects on the Majorana Fermi surfaces~\cite{Hermanns2}.

\bibliography{suppl}